\documentclass[9.5pt]{article}
\usepackage{spconf,amsmath,graphicx}
\usepackage{booktabs}
\usepackage{amsmath}
\usepackage{hyperref}
\usepackage{amssymb}
\usepackage{arabtex}
\usepackage{enumitem}
\usepackage{xcolor}
\usepackage{color, colortbl}
\usepackage{tabularx}
\usepackage{caption}
\usepackage{arabtex}
\usepackage{utf8}
\usepackage{graphicx}
\usepackage{multirow}
\usepackage{cite}
\usepackage{amsmath,amssymb,amsfonts}
\usepackage{algorithmic}
\usepackage{url}
\usepackage{caption}
\usepackage{subcaption}
\usepackage{color,soul}
\usepackage{booktabs}
\usepackage{multirow}
\usepackage{array}
\usepackage{balance}
\usepackage{lipsum}
\setcode{utf8}

\name{Yassine El Kheir, Ahmed Ali, Shammur Absar Chowdhury}
\address{Qatar Computing Research Institute, HBKU, Doha, Qatar}

\title{Speech Representation Analysis \\based on Inter- and Intra-Model Similarities }

\begin{document}
%
\maketitle
\begin{abstract}

Self-supervised models have revolutionized speech processing, achieving new levels of performance in a wide variety of tasks with limited resources. However, the inner workings of these models are still opaque. In this paper, we aim to analyze the encoded contextual representation of these foundation models based on their inter- and intra-model similarity, independent of any external annotation and task-specific constraint. We examine different SSL models varying their training paradigm -- Contrastive (Wav2Vec2.0) and Predictive models (HuBERT); and model sizes (base and large). 
We explore these models on different levels of localization/distributivity of information including (i) individual neurons; (ii) layer representation; (iii) attention weights and (iv) compare the representations with their finetuned counterparts.
Our results highlight that these models converge to similar representation subspaces but not to similar neuron-localized concepts\footnote{A concept represents a coherent fragment of knowledge, such as ``a class containing
certain objects as elements, where the objects have
certain properties''\cite{stock2010concepts}}. 
We made the code publicly available 
for facilitating further research, we publicly released our code\footnote{\href{https://github.com/QCRIVoice/XSSL_speech}.{https://github.com/QCRIVoice/XSSL\_speech}
}.

\end{abstract}
\vspace{-0.2cm}

\vspace{-0.3cm}

\section{Introduction}

Self-supervised Speech models like Wav2Vec2 \cite{baevski2020wav2vec} and HuBERT \cite{hsu2021hubert} have shown remarkable advancements in a variety of speech processing tasks, including speech recognition, emotion recognition, speaker verification, and language identification \cite{mohamed2022self, shon2022slue, borgholt2022brief} among others. 
This significant advancement over supervised state-of-the-art methods and the opaqueness of these models has sparked interest in understanding and exploring their internal mechanisms. 

Several studies have aimed to understand the information these models capture about different properties such as speaker characteristics \cite{fan2020exploring, chowdhury2023end,chowdhury2020does, feng2022silence}, paralinguistic aspects \cite{shah2021all, li2023exploration}, articulatory features \cite{ji2022predicting}, acoustic-linguistic elements \cite{pasad2021layer}, as well as accent features \cite{yang2023can} among others. Moreover, studies like \cite{pasad2021layer, pasad2023comparative} have also shown how better model understanding can lead to efficient fine-tuning strategies for downstream tasks.


A widely used interpretation technique includes training supervised classifiers, aka probing classifiers \cite{9414321,chowdhury2023end, belinkov2019analysis}, based on the learned representations of given models, to predict various task properties. This methodology has found application in various studies and showed the ability of representations from different models to capture distinct properties. Additionally, similarity-based methods are used to find associations at the frame-, phoneme-, and word-levels.
These methods utilizes metrics such as projected-weighted canonical correlation analysis ($pwcca$) \cite{morcos2018insights} and mutual information without training classifiers. However, the effectiveness of this approach is limited by the need for annotated data, requiring precise boundary alignment, accurate phoneme transcription, coupled with consistent word alignment, to ensure valid and reliable analysis and results.



In this study, we introduce inter- and intra-model similarity measures to understand contextual representations within speech models. Instead of focusing on a specific category or property of information, we focus on exploring both inter- and intra-similarity across a spectrum of models. We investigate localization/distributivity\footnote{Does every single neuron encode a single concept or all concepts are spread across multiple neurons? \cite{mcclelland1986appeal}} properties in these models. We adopted a set of $5$ distinct similarity measures, to explore the SSL models for localization/distributivity behavior at individual neurons, layers, and attention mechanisms levels. This comprehensive range of metrics allows us to capture the nuances in the patterns embedded within the models, offering a granular view of their structural and functional dynamics. 


Our in-depth analysis reaffirms prior discoveries without the need for external data or defined tasks. Moreover, our findings also reveal noteworthy insights: (i) Speech SSL model neurons exhibit higher intra-model similarity than inter-model similarity. (ii) Information encapsulated by neurons from one layer can be represented as a linear combination of other layers. Models have similar representation subspaces but different localized neuron concepts. (iii) Lower and adjacent layers demonstrate a high degree of similarity across diverse models. (iv) The training objective has a greater impact on representation similarity than the size of the model architecture. (v) Finally, we show how the similarity analysis can motivate efficient finetuning for ASR, where freezing the bottom layers of models still maintains comparable performance to finetuning the full network while reducing the finetuning time.

\begin{figure*}[!ht]
  \centering
  \includegraphics[width=1\textwidth]{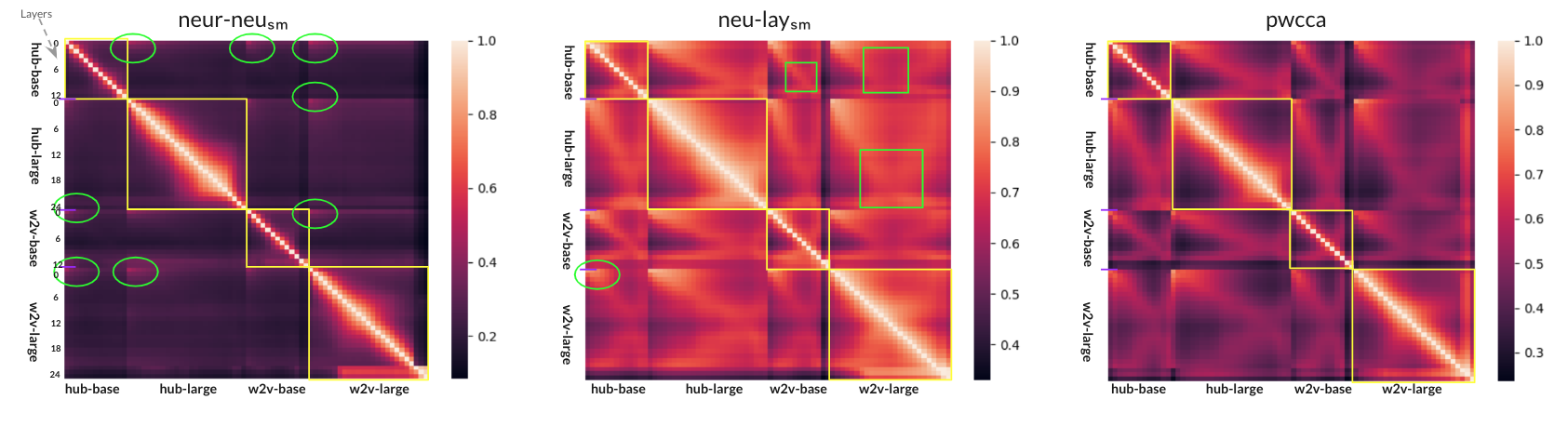}
  \vspace{-0.5cm}
  \caption{Comparison of Heatmap Similarities Between HuBERT and Wav2Vec2.0 Models: $\text{neu-neu}_{\text{sm}}$, $\text{neu-lay}_{\text{sm}}$, and $pwcca$ Similarities. Model boundaries are highlighted in yellow, and noteworthy similarities are encircled in green.}
  \label{fig:heatmaps_1}
\end{figure*}

\vspace{-0.3cm}

\section{Methodology}
\vspace{-0.25cm}
We analyzed $M$ pretrained speech SSL models for both localized and distributed information using various widely accepted similarity metrics \cite{wu-etal-2020-similarity}, capturing different notations at individual neurons, layers, and attention levels. We propose to remove any dependency on external labels or boundary annotation by utilizing only frame-level representation for the study.

For each model $m$, ($m \in M$), we extracted frame level representations $h^{(m)}_l$ $\in \mathbb{R}^{d_m}$,  where $d_m = \{768, 1024\}$, is indicative of number of neurons, and attention weights $\alpha^{(m)}_l$ at each layer $l$. 
We then exploit the extracted neuron/layer-level representation and attention weight to find inter- and intra-similarities at various levels of granularities.





\vspace{-0.3cm}

\subsection{Neuron-level Similarity}
\vspace{-0.25cm}
\label{sec:1}
We adopted two different similarity measures: \textit{(i)} neuron-neuron similarity ($\text{neu-neu}_{\text{sm}}$) --  similarity between pairs of individual neurons, and \textit{(ii)} neuron-layer similarity ($\text{neu-lay}_{\text{sm}}$) -- similarity between a neuron in one model with a layer in another.


For a given neuron $h^{(m)}_l[k]$ of a model $m$, and a layer $l$, $\text{neu-neu}_{\text{sm}}$ is defined as the maximum correlation between $h^{(m)}_l[k]$ and another neuron $h^{(m')}_l[k']$ of layer $l'$ of another model $m'$: 
\begin{equation}
\begin{aligned}
    \widetilde{\text{neu-neu}_{\text{sm}}}(h^{(m)}_l[k], h^{(m')}_{l'}) &=
    \max_{k'} \rho(h^{(m')}_{l'}[k'], h^{(m)}_l[k])
\end{aligned}
\end{equation}

\noindent Then, we average over all neurons in layer $l$ of the model $m$,
\begin{equation} 
\begin{aligned}
\text{neu-neu}_{\text{sm}}(h^{(m)}_l, h^{(m')}_{l'}) &=\\
\frac{1}{d_m} \times \sum_{k} \widetilde{\text{neu-neu}_{\text{sm}}}(h^{(m)}_l[k], h^{(m')}_{l'}).
\end{aligned}
\end{equation}
\noindent where $\rho$ is the Pearson correlation. 

\noindent $\text{neu-neu}_{\text{sm}}$ is designed to assess the \textbf{localization of information}, reflecting higher values when two layers exhibit pairs of neurons that demonstrate similar behavioral patterns.

\noindent In contrast, $\text{neu-neu}_{\text{sm}}$ assesses how can a neuron $h^{(m)}_l[k]$ be expressed as linear regression of neurons of another layer $l'$ of another model $m'$, and measures the quality of regression fit which is defined as: 

\begin{equation}
\begin{aligned}
\widetilde{\text{neu-lay}_{\text{sm}}}(h^{(m)}_l[k], h^{(m')}_{l'}) &= lstsq(h^{(m')}_{l'}, h^{(m)}_l[k]).r
\end{aligned}
\end{equation}

\noindent $lstsq$ denotes linear least-squares, and $r$ represents the associated r-value. As before, this is extended to the layer level:

\begin{equation}
\begin{aligned}
\text{neu-lay}_{\text{sm}}(h^{(m)}_l, h^{(m')}_{l'}) &= \frac{1}{d_m} \sum_{k} \widetilde{\text{neu-lay}_{\text{sm}}}(h^{(m)}_l[k], h^{(m')}_{l'})
\end{aligned}
\end{equation}

\noindent $\text{neu-lay}_{\text{sm}}$ reflects how the \textbf{localized information} in a particular neurons in $m$ are \textbf{distributed} across the the layers of models $m'$.
\vspace{-0.25cm}

\subsection{Representation-level Similarity}
\vspace{-0.25cm}
For layer-level representation analysis, we focus on canonical correlation analysis (CCA) similarities measures. Despite previous work that focuses on using projected-weighted CCA similarity ($pwcca$)\cite{pasad2021layer, pasad2023comparative} and singular vector CCA ($svcca$)\cite{raghu2017svcca}, we focus on examining similarities among frame representations instead of frame representation with other information such as phonemes, words, and boundaries, which typically require extensive annotation and linguistic expertise. These similarity measures underscore \textbf{the distributive nature of information across layers} which highlights scenarios where two layers exhibit similar behaviors across all their neurons, emphasizing the collective patterns rather than relying solely on individual neuron matching. 
\vspace{-0.25cm}

\subsection{Attention-level Similarity}
\vspace{-0.25cm}

Similar to $\text{neu-neu}_{\text{sm}}$, $\text{attention}_{\text{sm}}$ similarity identifies the most ``correlated" other attention heads within the model $m$ and across $m'$. This measure captures \textbf{the behavior similarity} indicating the focus alignment. 
Given two attention heads, $\alpha_{l}^m[k]$ and $\alpha_{l'}^{m'}[k']$, we calculate their similarity based on their Pearson correlation, then we average over the heads in layer $l$ as in Section \ref{sec:1}.



\vspace{-0.3cm}

\section{Experimental Setup}
\vspace{-0.25cm}

\paragraph*{SSL Models} We adopt widely used self-supervised speech models, HuBERT (hub) and Wav2Vec2.0 (w2v)\footnote{Available here: \href{https://huggingface.co/collections/facebook}{https://huggingface.co/collections/facebook}} as reported in Table \ref{tab:models_comparison}. Both models share similar architectures. The encoder network consists of blocks of temporal convolution layers with $512$ channels, and the convolutions in each block have strides and kernel sizes that compress about $25$ms of $16$kHz audio every $20$ms. The context network consists of ${12 \text{ (base)}}$ \text{and $24$ (large)} blocks with model dimension ${768 \text{ (base)}}$ \text{and $1024$ (large)} and attention heads of ${12 \text{ (base)}}$ \text{and $16$ (large)}. The underlying difference in the models lies in their training objectives; w2v undergoes training through Contrastive Predictive Coding (CPC) loss, employing masking techniques, thereby classifying it as a contrastive model. On the other hand, hub, follows a different approach by attempting to predict discrete targets of masked regions using Cross-Entropy (CE) loss, classifying the model as a predictive model.

\vspace{-0.25cm}

\paragraph*{Dataset} We use the extensively employed TIMIT dataset \cite{garofolo1993timit} in research, studies on phone recognition, phone segmentation, and speaker recognition. TIMIT comprises $5.4$ hours of clean data manually transcribed. Despite its limited size, the dataset features a diverse set of approximately $630$ speakers delivering phonetically rich sentences, rendering it favorable for our task. For our task focuses on studying similarities, we exclusively utilize the official training set. Given that we employ frame-level embeddings in this context, each of the selected models yields over $700K$ embeddings from each layer (each $20$ ms corresponds to an embedding).

\vspace{-0.25cm}

\begin{table}
\vspace{-0.25cm}

\centering
\resizebox{0.4\textwidth}{!}{%
\begin{tabular}{lcc}
\textbf{Models} & (\textit{Abbreviation}) & \textbf{Training Data} \\
\hline
\hline
HuBERT (BASE) & hub-base & Librispeech 960hrs           \\
\hline
HuBERT (LARGE) & hub-large & Libri-Light               \\
\hline
Wav2Vec 2.0 (BASE) & w2v-base & Librispeech 960hrs          \\
\hline
Wav2Vec 2.0 (LARGE) & w2v-large & Libri-Light            
\end{tabular}%
}
\caption{Examined Pretrained Speech SSL Models}
\label{tab:models_comparison}
\vspace{-0.5cm}

\end{table}

\section{Analysis \& Discussion}


\paragraph*{A. Neuron Intra-Model Similarity.} 
\label{intra}
\vspace{-0.25cm}

Figure \ref{fig:heatmaps_1} illustrates heatmaps showing similarities between different neurons and layers across various models using $\text{neu-neu}_{\text{sm}}$, $\text{neu-lay}_{\text{sm}}$, and $pwcca$ similarities. The $\text{neu-neu}_{\text{sm}}$ reveals a distinctive diagonal pattern within each model, indicating that neurons within a specific model $m$ and layer $l$ tend to exhibit similarity to their counterparts neurons in adjacent layers within the same model $m$. However, individual neurons are very different when comparing a given model $m$ individual neurons to other models neurons. This observation suggests that neurons exhibit significantly higher \textit{intra-model similarities} than \textit{inter-model similarities}. A similar pattern was found in contextualized language models \cite{wu-etal-2020-similarity}. Furthermore, the identified similarity pattern in lower layer neurons is consistent across all examined models, potentially linked to their proximity to CNN feature extraction layers equivalent to spectrogram features, as demonstrated in previous work \cite{wu2022performance}.


\vspace{-0.25cm}



\begin{figure}[!ht]
  \centering
  \includegraphics[width=0.4\textwidth]{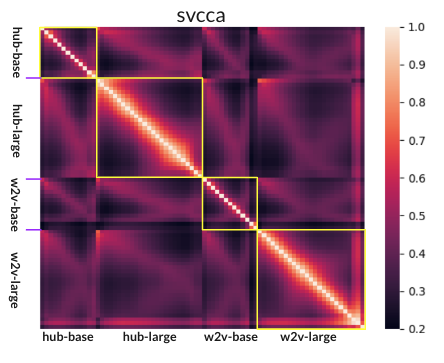}
  \vspace{-0.3cm}
  \caption{Heatmap of $svcca$ similarity}
 \vspace{-0.2cm}
  \label{fig:heatmaps_svcca}
\vspace{-0.45cm}

\end{figure}

\begin{figure*}[!ht]
  \centering
  \includegraphics[width=1\textwidth]{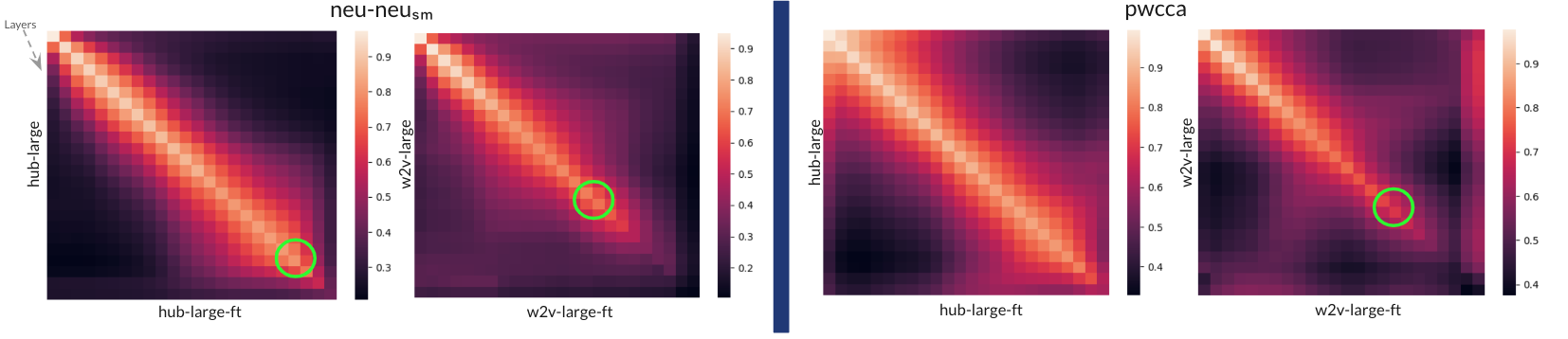}
  \vspace{-0.7cm}
  \caption{Comparing Heatmap Similarities between HuBERT and Wav2Vec2.0 Models in Relation to their ASR Finetuning Variations: $\text{neu-neu}_{\text{sm}}$ and $pwcca$ Similarities. Noteworthy similarities are encircled in green.}
  \label{fig:heatmaps_3}
  \vspace{-0.25cm}

\end{figure*}
\vspace{-0.25cm}
\paragraph*{B. Layer Inter-Model Similarity.}
While individual pairs exhibit distinct characteristics across different models, $\text{neu-lay}_{\text{sm}}$ in Figure \ref{fig:heatmaps_1} reveals strong
inter-model similarity, suggesting that the representations of different models converge to similar subspaces. Furthermore, the results also suggest that the individual neurons of a model can be represented as a linear combination of neurons from other layers of the model. These cross-model similarities are also observed using representation-level similarities $pwcca$, and $svcca$.

\vspace{-0.5cm}
\paragraph*{C. Models within the same family behaves similarly.}


Notably, $\text{neu-lay}_{\text{sm}}$ similarity reveals that neurons concepts in the top layers of hub based models (base and large) are less similar to lower layers, and vice versa, which supports the fact that the lower layer captures different fine-grained concepts, whereas the higher layers are capturing more abstract information as seen in \cite{pasad2021layer, pasad2023comparative, chowdhury2023end}. 
In contrast, w2v models show a different trend. A notable similarity is observed in the higher layers of both base (layer L8 - L10) and large (L20 - L23) models with respect to all the layers within the model. These intra-model similarities are seen using both $pwcca$ and $svcca$ similarity measures.
Despite the high similarities in inter-model layer representation (as shown in Section 4.B), the final layer of the hub and w2v (both base and large) are very distinct. Our observations indicate that the models within the same family (base and large) exhibit behavioral similarities in representation. We hypothesize the uniqueness in representation between the family -- hub {\em vs} w2v is more likely attributed to the training objective of self-supervised models rather than the architecture's number of layers which aligns with the findings reported in \cite{9414321}.

\vspace{-0.25cm}
\paragraph*{D. Adjacent and Lower Layers Similarity.} All the heatmaps in Figure \ref{fig:heatmaps_1}, including the $svcca$ similarity in Figure \ref{fig:heatmaps_svcca}, show a bright diagonal and its neighboring areas. This brightness suggests that neighboring layers share similar representations, indicating that adjacent layers in the models encapsulate similar information subspaces. Similar patterns are observed in both language models and vision networks \cite{kornblith2019similarity}. Additionally, $pwcca$ discloses that lower layers subspaces demonstrate similarity across studied models. This alignment with expectations, and with previous findings in Section 4.A, where lower layers closely resemble CNN layers functioning as feature extractors, and these features exhibit equivalence representations across the considered models.

\vspace{-0.3cm}

\begin{figure}[!ht]
  \centering
  \includegraphics[width=0.4\textwidth]{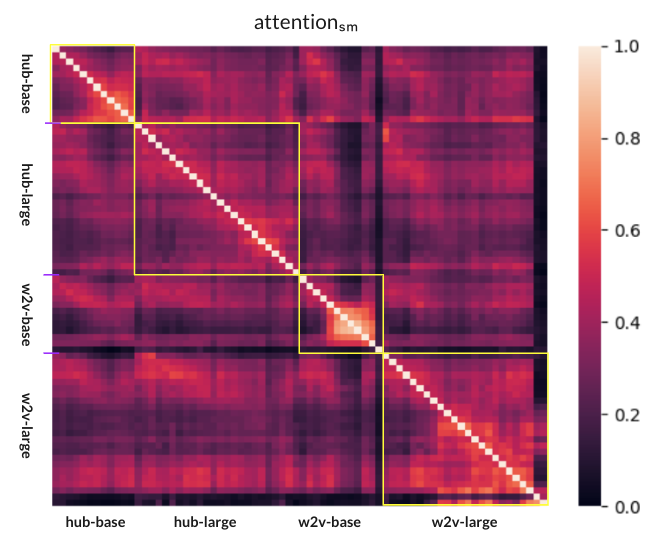}
\vspace{-0.25cm}
  \caption{Heatmap of $\text{attention}_{\text{sm}}$ similarity}
  \label{fig:heatmaps_attention}
  \vspace{-0.5cm}

\end{figure}

\paragraph*{E. Attention Weights Similarity.} Examining Figure \ref{fig:heatmaps_attention}, we observe 
high similarities between the attention heads in the upper layers of the models with respect to the lower layers. These high similarities in the upper layer could indicate redundancy in design. However, it is important to note that attention-based similarity measures are not reliable and are harder to interpret, as fine-grained patterns are less noticeable in the similarity-based analysis. 



\vspace{-0.25cm}

\paragraph*{F. ASR Finetuning Effect.} Figure \ref{fig:heatmaps_3} depicts the similarities in heatmaps between w2v-large, and hub-large, along with their ASR fine-tuned counterparts w2v-large-ft and hub-large-ft, on Librispeech dataset. The analysis utilizes $pwcca$ and $\text{neu-neu}_{\text{sm}}$ similarities to explore potential changes in information across different layers and the localized information within neurons. Results indicate that hub-based models primarily undergo significant changes after fine-tuning only at the last few layers in comparison to other layers (similarity between the foundation and its finetuned counterpart is less than $0.5$ in upper layers). For w2v model, we observe that a large number of upper layers has changed significantly at both the neuron and layer levels, in alignment with the findings in \cite{pasad2021layer}. Such findings indicate that finetuning exclusively the upper layers can be as effective as finetuning the full model. Our observation is aligned with the findings in \cite{pasad2023comparative} where this conclusion was drawn by examining phoneme-level $cca$ following the fine-tuning of only the layer $16$ in w2v-large and layer $20$ in hub-large, which yielded comparable results to finetuning all parameters. Note in \cite{pasad2023comparative} used human annotation for phoneme boundaries, whereas our proposed method gave the same conclusion without relying on any external annotation.
\vspace{-0.5cm}

\paragraph*{Key Points.} Our study highlights how different models trained with distinct objectives can converge toward similar representations and concepts. We observed that neurons in one layer can be expressed as linear combinations of neurons from other layers in different models. Importantly, this convergence is driven more by the distributivity nature of representations than by neuron concept localization. In other words, individual neurons learn different localized concepts, but overall, they contribute to similar subspaces across layers. 
\vspace{-0.4cm}

\section{Conclusion}
\vspace{-0.25cm}

The paper introduces both annotation- and task-independent approaches for analyzing various speech SSL models. Our in-depth analysis explores both Wav2Vec2.0 and HuBERT model families, revealing intricate convergence patterns in inter- and intra-model neurons, layers, and attention weights similarities. Our finding suggests that models share similar distributional representations but different localized concepts, and the training objective emerges as a pivotal factor, outweighing the influence of model size. Hence, signaling how understanding the inner workings of such large models can facilitate effective and parameter-efficient design decisions for both foundation and downstream models.

\bibliographystyle{IEEEbib}
\bibliography{main}

\end{document}